\documentclass[prl,graphicx,reprint,twocolumn, superscriptaddress]{revtex4}
\usepackage{graphics,bm,overpic,subfigure,color,amsmath}

\newcommand{\figref}[1]{Fig.~\ref{#1}}
\newcommand{\figrefs}[1]{Figs.~\ref{#1}}

\newcommand{\rudolphpeierls}{Rudolph Peierls Centre for Theoretical Physics, University of Oxford, 1 Keble Road,  Oxford, OX1 3NP, UK}
\newcommand{\blackett}{Blackett Laboratory, Imperial College, London, SW7 2AZ, UK}
\newcommand{\culham}{EURATOM/CCFE Fusion Association, Culham Science Centre, Abingdon, OX14 3DB, UK}

\newcommand{\magdalencollege}{Magdalen College, Oxford, OX1 4AU, UK}

\newcommand{\mitpsfc}{Plasma Science and Fusion Center, MIT, Cambridge, MA 02139, USA}
\newcommand{\maryland}{Department of Physics, University of Maryland, College Park, MD 20742, USA}

\newcommand{\edmundhighcock}{
\author{E.\ G.\ Highcock}
\email{edmund.highcock@physics.ox.ac.uk}
\affiliation{\magdalencollege
}
\affiliation{\rudolphpeierls
}
\affiliation{\culham
}

}

\newcommand{\michaelbarnes}{
\author{M.\ Barnes}
\affiliation{\mitpsfc
}

}

\newcommand{\felixparra}{
\author{F.\ I.\ Parra}
\affiliation{\mitpsfc
}

}

\newcommand{\colinroach}{
\author{C. M. Roach}
\affiliation{\culham
}

}

\newcommand{\stevecowley}{
\author{S. C. Cowley}
\affiliation{\culham
}
\affiliation{\blackett
}

}

\newcommand{\alexschekochihin}{
\author{A.\ A.\ Schekochihin}
\affiliation{\rudolphpeierls
}

}

\newcommand{\billdorland}{
\author{W.\ Dorland}
\affiliation{\maryland
}
}

\newcommand{\myfig}[3][3.3in]{
\begin{figure}[htbp]
\includegraphics[width=#1]{#2}%
\caption{#3\label{#2}}%
\end{figure}
}

\newcommand{\kapdefeq}{}
\newcommand{\kapcom}{R/L_T}
\newcommand{\kapcritcom}{R/L_{Tc}}

\begin{document}


\title{Zero-Turbulence Manifold in a Toroidal Plasma} 

\edmundhighcock
\alexschekochihin
\stevecowley
\michaelbarnes
\felixparra
\colinroach
\billdorland

\date{\today}

\begin{abstract}
Sheared toroidal flows can cause bifurcations to zero-turbulent-transport states in tokamak plasmas.
The maximum temperature gradients that can be reached are limited by subcritical turbulence driven by the parallel velocity gradient.
Here it is shown that $q/\epsilon $ (magnetic field pitch/inverse aspect ratio) is a critical control parameter for sheared tokamak turbulence.
By reducing $q/\epsilon$, far higher temperature gradients
can be achieved without triggering turbulence,
in some instances comparable to those found experimentally in transport barriers. 
The zero-turbulence manifold
is mapped out,
in the zero-magnetic-shear limit, over
the parameter space ($\gamma_E $, $q/\epsilon $, $R/L_T $),
where $\gamma_E $ is the perpendicular flow shear and
$R/L_T$ is the normalised inverse temperature gradient scale.
The extent to which it can be constructed from linear theory is discussed.
\end{abstract}


\maketitle 


\paragraph{Introduction.}
The heat loss that occurs as a result of turbulence
driven by the ion temperature gradient (ITG)
is one of the main obstacles to a successful fusion reactor.
A large body of experimental work has demonstrated
the effectiveness of strongly sheared equilibrium-scale flows
in reducing this turbulence \cite{burrell1997effects,wolf2003internal,vries2009internal}.
Numerical models \cite{waltz1994toroidal,dimits2001parameter,kinsey2005flowshear,roach2009gss} have
demonstrated that by reducing the strength of the
 ITG instability, which drives the turbulence, and by shearing apart the turbulent structures, the radial gradient of the flow component perpendicular to the magnetic field can indeed lead to a great reduction in the heat loss that results from a given temperature gradient.
However, Ref. \cite{kinsey2005flowshear} also demonstrated that the instability associated with the parallel velocity gradient (PVG)
\cite{catto1973parallel,newton2010understanding,schekochihin2010subcritical}
could start to drive turbulence at higher flow gradients and prevent the complete suppression of the turbulent transport.

More recent work
has demonstrated that, at even higher flow shears,
it is possible,
at moderate temperature gradients,
for the perpendicular velocity shear to overcome both the ITG and the PVG instabilities and completely quench the turbulence \cite{barnes2011turbulent}.
This quenching is most effective at zero magnetic shear
\cite{highcock2010transport,highcock2011transport,parra2010plausible},
a regime which has been associated in experiments with high confinement of energy
in the presence of strongly sheared flows \cite{sips1998operation,vries2009internal}.
Refs. \cite{highcock2010transport,highcock2011transport,parra2010plausible}  also demonstrated the existence, at zero magnetic shear, of a bifurcation to a high-temperature-gradient reduced-transport state, driven by a toroidal sheared flow.
However, the maximum temperature gradient that could be reached via
such a bifurcation was found to be limited by the fact that turbulence was rekindled at high toroidal shear, in the form of subcritical fluctuations driven by the PVG \cite{newton2010understanding,schekochihin2010subcritical,barnes2011turbulent,highcock2010transport,highcock2011transport}.
The question arises,
which parameter regime is most favourable to the suppressing effect of the perpendicular flow shear and least favourable to the ITG and PVG drives? 
 In other words, how can the temperature gradient which results from the transport bifurcation described in Refs. \cite{highcock2010transport,highcock2011transport,parra2010plausible} be maximised?

 At zero magnetic shear, the turbulence is subcritical for all nonzero values of the flow shear: there are no linearly unstable eigenmodes,
 and sustained turbulence is the result of nonlinear interaction between linear modes which grow only transiently before decaying.
 A recent paper \cite{schekochihin2010subcritical},
which studied this transient growth in slab geometry,
demonstrated that at large velocity shears the maximal amplification 
exponent of a transiently growing perturbation before it decays is proportional
to the ratio of the PVG to the perpendicular flow shear.
In a torus, this quantity is equal to the ratio of the toroidal to poloidal magnetic field components, or $q/\epsilon $, where $q $ is the magnetic safety factor and $\epsilon $ is the inverse aspect ratio.
Therefore, if we conjecture that a certain minimum amplification exponent is required for sustained turbulence, Ref. \cite{schekochihin2010subcritical} predicts that there should be a value of $q/\epsilon $ below which the PVG drive is rendered harmless.
Below that value of $q/\epsilon $, it should be possible to maintain an arbitrarily high temperature gradient
without triggering turbulent transport provided a high enough perpendicular flow shear can be achieved.

In this Letter, motivated by the possibility of reduced transport at low values of $q/\epsilon$,
we use nonlinear gyrokinetic simulations to map out \emph{the zero-turbulence manifold},
the surface in the parameter space that divides the regions where turbulent transport can and cannot be sustained.
The parameter space we consider is ($\gamma_E $, $q/\epsilon $, $\kapcom $), where $\gamma_E $ is the normalised perpendicular flow shear: 
$	\gamma_E= u'/(q/\epsilon)$, where
$u'= dR\omega/dr/(v_{thi}/R)$
is the toroidal shear, $\omega $ the toroidal angular velocity, $r $ the minor radius of the flux surface, $v_{thi} $ the ion thermal speed and $R$ the major radius, and where $\kapdefeq R/L_T $ is the inverse temperature gradient scale length normalised to $R$. For brevity, we will  refer to $R/L_T$ as ``the temperature gradient.''
We set the magnetic shear to zero, the regime we expect to be most amenable to turbulence quenching by shear flow \cite{barnes2011turbulent, highcock2010transport, highcock2011transport, parra2010plausible, vries2009internal, sips1998operation, mantica2011ion}.

We discover that reducing $q/\epsilon $ is indeed uniformly beneficial to maintaining high temperature gradients in a turbulence-free regime,
and that values of $\kapcom$ can be achieved
that are comparable to those experimentally observed for internal transport barriers \cite{vries2009internal,field2011plasma}.

In the next sections, having presented our numerical model and methodology, we will describe these results and discuss their physical underpinnings, as well as their implications for confinement in a toroidal plasma. We will show that linear theory of subcritical fluctuations \cite{schekochihin2010subcritical} can, with certain additional assumptions, provide good predictions of the nonlinear results.

\paragraph{Numerical Model.}

To model the turbulence, we use the gyrokinetic equation \cite{frieman1982nge} in the high-flow, low-Mach limit \cite{flowtome1} (i.e., the toroidal rotation velocity is ordered to be smaller than the sound speed but much larger than the diamagnetic velocity; Coriolis and centrifugal 
effects are neglected \footnote{
The impact of these effects on turbulence is studied in Refs.~\cite{peeters2007toroidal,casson2010gyrokinetic}.
}, but velocity gradients are retained).
We take the electrostatic limit and assume a modified Boltzmann electron response.
The model used is identical to that in Ref. \cite{highcock2011transport}.
The gyrokinetic
system of equations is solved using the local nonlinear simulation code \texttt{GS2} \cite{gs2ref, jenko:1904, gyrokineticswebsite}.
As in Ref. \cite{highcock2011transport},
we take the Cyclone Base Case parameter regime \cite{dimits:969},
i.e., concentric circular flux surfaces with $\epsilon = 0.18 $,
inverse ion density scale length
 $R / L_{n}=2.2 $
 and ion to electron temperature ratio $T_i/T_e=1$.\footnote{
 A temperature ratio of 1 is appropriate for both lower power and future reactor-like conditions, but not for high-performance shots in current devices \cite{mantica2011ion,team1999alpha,petty1999dependence}.
 }
The magnetic shear is  $\hat{s}=0$. 
The ratio $q/\epsilon $ is varied by varying $q $ alone. Collisions are included by means of a model collision operator, which includes scattering in both pitch angle and energy and which locally conserves energy, momentum and particles \cite{abel2008linearized, numerics}.
The resolution of all simulations was \(128\times128\times40\times28\times8\) (poloidal, radial, parallel, pitch angle, energy). 
Note that relatively high parallel resolution was needed to resolve the PVG modes \cite{highcock2011transport}.

\myfig{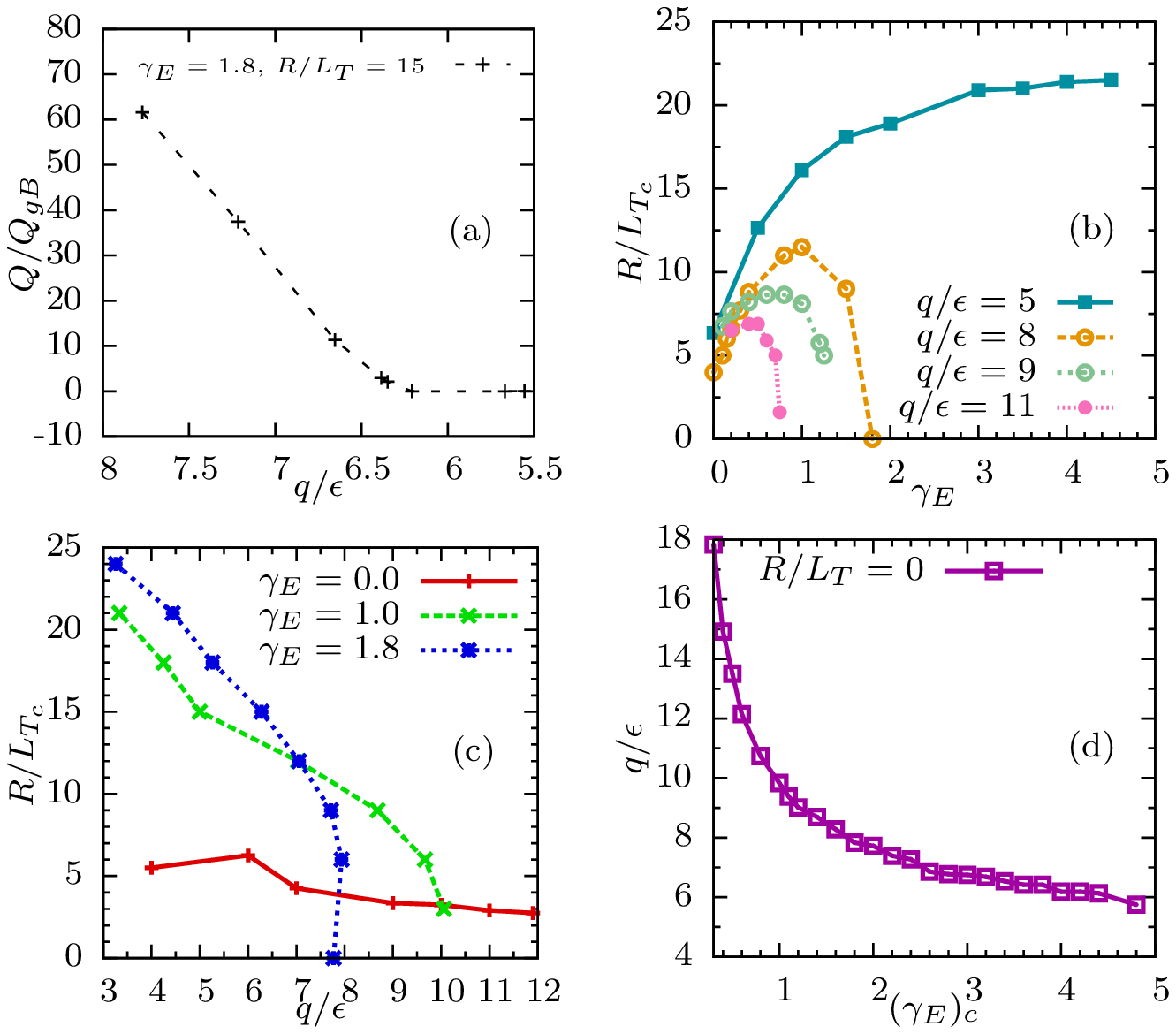}{(a) The simulations used to find the point on the manifold $\gamma_E=1.8$, $\kapcom=15$, $q/\epsilon=6.3$, showing the heat flux vs. $q/\epsilon$ at ($\gamma_E=1.8$, $\kapcom=15$). The point on the manifold is the point where the heat flux drops to zero.  (b-d) Sections through the critical manifold with parameters as indicated.
Turbulence cannot be sustained for $\kapcom<\kapcritcom$ in (b,c), or for $\gamma_E<\gamma_{Ec}$ in (d).
The data points were found as illustrated in (a), and used to generate the manifold shown in \figref{criticalcurve3dall}.
}
\paragraph{Method.}
We wish to determine, in a three-dimensional parameter space ($\gamma_E$, $q/\epsilon$, $\kapcom$),  the boundary  between the regions where turbulence can and cannot be sustained nonlinearly.
We cover this space using four scans with constant $q/\epsilon $  (\figref{projections}(b)), three scans with constant $\gamma_E$ (\figref{projections}(c)) and one scan with  constant $\kapcom$ (\figref{projections}(d)).
For each of these cases, we consider multiple values of a second parameter and
find the value of the third parameter corresponding to the zero-turbulence boundary.
The boundary is defined as the point where both the turbulent heat flux and the turbulent momentum flux vanish.
Thus, the location of each single point on the boundary is determined using on the order of ten nonlinear simulations.
An example of this procedure is shown in \figref{projections}(a).
In total, we performed more than 1500 simulations to produce the results reported below.

Because the turbulence that we are considering is subcritical, there is always a danger that a simulation might fail to exhibit  a turbulent stationary state because of an insufficient initial amplitude
\cite{baggett1995mostly,kerswell2005recent}.
As we are not here concerned with the question of critical initial amplitudes
we will consider a given set of parameters to correspond to a turbulent state if such a state can be sustained starting with a large enough perturbation.
Therefore, all simulations are initialised with high-amplitude noise.
They are then run to saturation; close to the boundary a simulation may need to run for up to $t\sim 1000 R/v_{thi}$ to achieve this.

The critical curves obtained in this manner are plotted in \figrefs{projections}(b-d).
These curves, which effectively give the critical temperature gradient $\kapcritcom$ as a function of $\gamma_E$ and $q/\epsilon$,
are then used to interpolate a surface, the zero-turbulence manifold, plotted in \figref{criticalcurve3dall}.
The interpolation is carried out using radial basis functions with a linear kernel \cite{buhmann2001radial} (see also \cite{highcock2011transport}).

\myfig{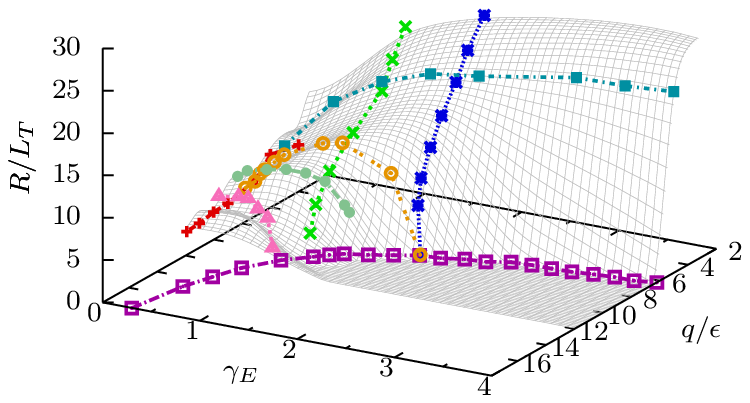}{ The zero-turbulence manifold. Turbulence can be sustained at all points outside the manifold (that is, at all points with a higher temperature gradient and/or higher value of $q/\epsilon$ than the nearest point on the manifold).  This plot is made up from the sections shown in \figref{projections}(b-d) (heavy lines) and the manifold interpolated from them (thin grey mesh).}
\myfig{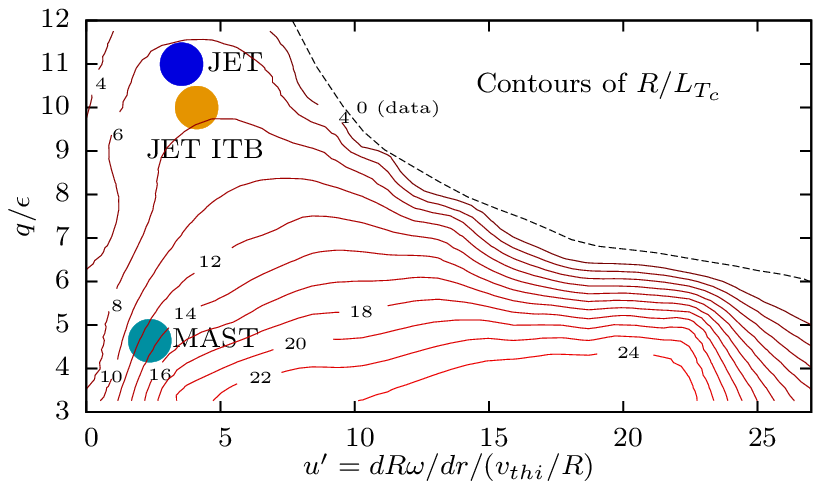}{Contours of the zero-turbulence manifold plotted against the toroidal flow shear $u'=dR\omega/dr / (v_{thi}/R)=\gamma_E/(q/\epsilon)$. The contours indicate the value $\kapcom=\kapcritcom$ below which turbulence is quenched.
From top to bottom, the circles indicate approximate values of $u'$ and $q/\epsilon$, corresponding to Ref. \cite{mantica2011ion} (JET; $R/L_{T}\sim 8$), Ref. \cite{vries2009internal} (JET ITB; $R/L_{T}\sim 17$; note large discrepancy, see text) and Ref. \cite{field2011plasma} (MAST ITB; $R/L_T\sim 10$).
}

\paragraph{Results.} The results of the scan described above are displayed in \figrefs{projections}(b-d). These three figures show, at fixed values of either $\gamma_E $, $\kapcom $ or $q/\epsilon $, the threshold in either $\kapcom $ or $q/\epsilon $ below which turbulence cannot be sustained; they are, in effect, sections through the zero-turbulence manifold.

Considering first \figref{projections}(b), we see that, at fixed $q/\epsilon $, the critical gradient  $\kapcritcom $ first rises with $\gamma_E $, as the perpendicular flow shear suppresses the ITG-driven turbulence, and then falls --- in most cases to 0 --- as the PVG starts to drive turbulence instead. This phenomenon was discussed at length in \cite{barnes2011turbulent, highcock2010transport, highcock2011transport} (indeed the curve for $q/\epsilon=8 $ is taken from \cite{highcock2011transport}).
 Thus, for every $q/\epsilon $, there is an optimum value of the perpendicular flow shear $\gamma_E $
(and hence of the toroidal shear $u' $) for which the critical temperature gradient $\kapcritcom $ is maximised.
We see that reducing
$q/\epsilon$
increases
the maximum $\kapcritcom$ that can be achieved without igniting turbulence.
\figref{projections}(c) shows that this rule applies for all considered values of flow shear
\footnote{The increase at $\gamma_E=0 $ cannot, of course, be due to reduction of the PVG;
we assume that this occurs because of the simultaneous reduction of the maximum parallel length scale $qR$ in the system, leading to weaker ITG turbulence; see \cite{barnes2011critically}.}.
This is to be expected, because lower $q/\epsilon $ means weaker PVG relative to the perpendicular shear,
allowing higher values of the perpendicular flow shear to suppress the ITG before the PVG drive takes over.

Lastly, \figref{projections}(d) shows the threshold in $\gamma_E $ above which the PVG can drive turbulence alone, without the help of the ITG;
in other words, even configurations with a flat temperature profile would be unstable.
At very high $q/\epsilon $, already a very small flow shear will drive turbulence; as $q/\epsilon $ decreases, higher and higher values of $\gamma_E $ are required for the PVG turbulence to be sustained.
It cannot be conclusively determined from this graph whether, 
as suggested by linear theory \cite{schekochihin2010subcritical},
there is 
a finite critical value of $q/\epsilon $ below which PVG turbulence cannot be sustained, i.e., a nonzero value of $q/\epsilon $ corresponding to $\gamma_{Ec}\rightarrow \infty $.
However, for $q/\epsilon\lesssim 7  $, the critical $\gamma_E $ is far above what might be expected in an experiment \footnote{
By order of magnitude, $\gamma_E\sim M/q$, where $M$ is the Mach number of the toroidal flow. Thus, values of $\gamma_E$ much above unity are unlikely to be possible.
}, and so the $\gamma_E\rightarrow\infty$ limit is somewhat academic.
A definite conclusion we may draw is that at experimentally relevant values of shear, pure PVG-driven turbulence cannot be sustained for $q/\epsilon\lesssim 7$.

The zero-turbulence manifold interpolated from the numerical data points is displayed in \figref{criticalcurve3dall}.
The manifold comprises three main features: a ``wall'' where the critical temperature gradient increases dramatically at low $q/\epsilon $; a ``spur'' at low $\gamma_E $, jutting out to high $q/\epsilon $
(where, as $\gamma_E$ increases, the ITG-driven turbulence is suppressed somewhat before the PVG drive becomes dominant),
and finally the curve where the manifold intercepts the plane $\kapcom=0 $, whose shape is described above.

\paragraph{Practical Implications and Comparison with Experiment.} 
In order to illustrate better the implications of our findings for confinement,
we plot, in \figref{manifoldcontoursqedwdr}, contours of
$\kapcritcom$ versus $q/\epsilon $
and the toroidal flow shear $u' = dR\omega/dr/(v_{thi}/R) $.
The basic message is clear:
the lower the value of $q /\epsilon $, the higher the temperature gradient that can be achieved without igniting turbulence.
Once we have obtained the lowest possible value of $q/\epsilon $,
there is an optimum value of $u' $ which will lead to that maximum $\kapcritcom $.
We note that the dependence of this optimum value of $u' $ on $q/\epsilon $
is not as strong as the dependence of the optimum value of $\gamma_E$ on $q/\epsilon$
(clearly this must be so because $u'=(q/\epsilon)\gamma_E $).
In a device with an optimised value of $q/\epsilon $, a near maximum critical temperature gradient would be achievable for $u'\gtrsim 5$, shears comparable to those observed in experiment \cite{mantica2011ion,field2011plasma,vries2009internal}.


While simulation results obtained for Cyclone Base Case parameters are not suitable for detailed quantitative comparison with real tokamaks, it is appropriate to ask whether our results are at all compatible with experimental evidence.
For an internal transport barrier (ITB) in MAST, Ref. \cite{field2011plasma} reports $R/L_T\sim 10 $ at $q/\epsilon\sim 4.6 $ \footnote{
The ratio of toroidal to poloidal field in MAST can be smaller on the outboard side, so the effective value of $q/\epsilon$ for locating this case on the zero-turbulence manifold might be smaller than quoted.}
and $u'\sim 2.4 $. This is comparable to the critical values shown in \figref{manifoldcontoursqedwdr}. In JET, Ref. \cite{mantica2011ion} reports $R/L_T\sim8 $ at $q/\epsilon\sim 11 $ and $u'\sim 3.6 $, again reasonably close to what we would have predicted.
However, an ITB in JET studied by Ref. \cite{vries2009internal} achieved $R/L_T\sim17 $ at $q/\epsilon\sim10 $ and  $u'\sim4.1 $ --- substantially higher than our $R/L_{Tc} $ at the same values of $q/\epsilon $ and $u' $. Note, however, that Ref. \cite{vries2009internal}
reports that the shear was dominated by an enhanced poloidal flow, an effect which is not included in our numerical model.

\paragraph{Relation to Linear Theory.} Since the mapping of the zero-turbulence manifold using nonlinear simulations is computationally expensive, we may ask whether linear theory can predict marginal stability.
 The question is also interesting in terms of our theoretical understanding of subcritical plasma turbulence.
It is clear that in a situation where perturbations grow only transiently,
existing methods based on looking for marginal stability of the fastest growing eigenmode will not be applicable.
In Ref. \cite{schekochihin2010subcritical}, we considered these transiently growing modes in a sheared slab, and posited a new measure of the vigour of the transient growth: $N_{\mathrm{max}} $, the maximal amplification exponent, defined as the number of e-foldings of transient growth a perturbation experiences before starting to decay, maximised over all wavenumbers.
 It appears intuitively clear that in order for turbulence to be sustained,
 transient perturbations must interact nonlinearly before they start to decay.
 We may then assume that a saturated turbulent state will exist if $N_{\mathrm{max}}\gtrsim N_c $, where $N_c $ is some threshold value of order unity. The zero-turbulence manifold is then the surface $N_{\mathrm{max}}(\gamma_E,q/\epsilon,\kapcom)=N_c $.

We now test this idea by calculating $N_{\mathrm{max}} $ for linear ITG-PVG-driven transient perturbations in a slab, using the code \texttt{AstroGK} \cite{numata2010astrogk} to solve the linearised gyrokinetic equation, as done in Ref. \cite{schekochihin2010subcritical}. \figref{fit}(a) shows that for each value of $\gamma_E $ and a range of $q/\epsilon $, it is possible to choose $N_c(\gamma_E) $ such that the equation $N_{\mathrm{max}}(\gamma_E,q/\epsilon,\kapcom)=N_c $ correctly reproduces the critical curve $\kapcritcom(q/\epsilon) $ obtained as a section of the zero-turbulence manifold at that value of $\gamma_E $.
However, $N_c $ does have a strong dependence on $\gamma_E $, shown in \figref{fit}(b), ranging from $N_c\lesssim 0.5 $ at $\gamma_E \gtrsim 2 $ to $N_c \rightarrow \infty $ as $\gamma_E\rightarrow 0 $ (the latter is an expected result: at $\gamma_E=0 $, there is a growing eigenmode, so either $N_{\mathrm{max}}=\infty $ or there is no growth at all).
It is not clear if $N_c $ tends to a finite limit as $\gamma_E\rightarrow \infty $, but, similarly to the existence of a critical value of $q/\epsilon $ as $\gamma_E\rightarrow\infty $, this is a somewhat academic question because such a limit would be achieved (or not) at $\gamma_E $ too large to be experimentally achievable.

 The practical conclusion of this exercise is that
all that appears to be required to determine the two-dimensional dependence of $\kapcritcom $ on $\gamma_E $ and $q/\epsilon $ is finding $N_c(\gamma_E) $ using a nonlinear scan at a single value of $q/\epsilon $; thus, the number of parameters in the nonlinear scan is reduced by one.

\myfig[3.45in]{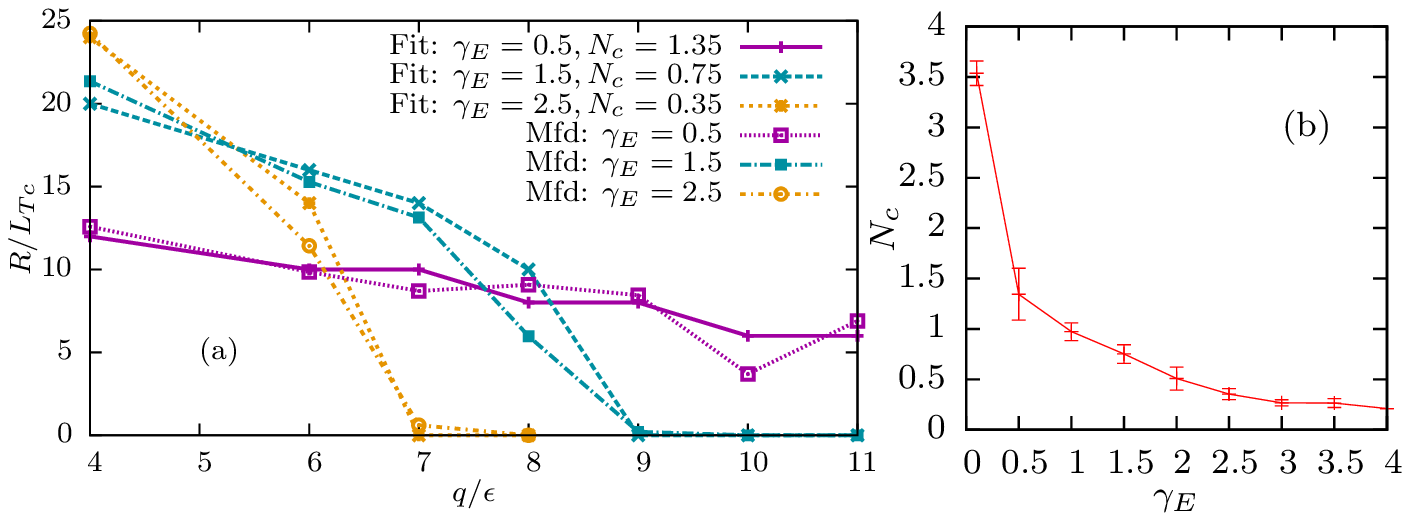}{(a) The critical temperature gradient $\kapcritcom$ vs. $q/\epsilon$ for different values of $\gamma_E$, showing both $\kapcritcom$ obtained from the interpolated manifold, and $\kapcritcom$ such that $N_{\mathrm{max}} = N_c$, with $N_c$ suitably chosen for each $\gamma_E$, as shown in (b).} 
\paragraph{Discussion.}  We have presented two key results.
Firstly, and principally, we have calculated
the shape of the zero-turbulence manifold, the surface that divides the regions in the parameter space ($\gamma_E $, $q/\epsilon $, $\kapcom $) where subcritical turbulence can and cannot be nonlinearly sustained.
We have described the shape of this manifold and its physical origins, and presented its two implications for confinement in toroidal plasmas: that reducing the ratio $q/\epsilon $,
i.e.,
increasing the ratio of the poloidal to the toroidal magnetic field,
improves confinement at every nonzero value of $\gamma_E $, and that at fixed $q/\epsilon $, there is an optimum value of $\gamma_E $ (that is, an optimum value of the toroidal flow shear $u'= dR\omega/dr/(v_{thi}/R) $) at which the critical temperature gradient is maximised,
 in some instances to values comparable 
to those observed in internal transport barriers \cite{vries2009internal,field2011plasma}.
How to calculate the heat and momentum fluxes that would need to be injected
in order for such optimal temperature gradients to be achieved
was discussed in Ref.~\cite{parra2010plausible}.

Secondly, we have shown that the zero-turbulence manifold can be parameterised as $N_{\mathrm{max}}(\gamma_E, q/\epsilon, \kapcom) = N_c(\gamma_E) $, where $N_{\mathrm{max}} $ is the maximal amplification exponent of linear transient perturbations (calculated from linear theory) and $N_c $ must be fit to the data.
Thus, using a single scan at constant $q/\epsilon $ to determine $N_c(\gamma_E) $ appears to be sufficient for calculating the full two-parameter dependence of the critical temperature gradient.
Obviously, the need to fit $N_c(\gamma_E) $ indicates a limitation of our current theoretical understanding of the criterion for sustaining subcritical turbulence in a sheared toroidal plasma.
The results reported here provide an empirical constraint on future theoretical investigations.

 Another avenue for future investigations is determining the dependence of the zero turbulence boundary on some of the parameters that were held fixed in this work:
$T_i/T_e$,
magnetic shear, and, more generally, the shape of the flux surfaces, density gradient, inverse aspect ratio $\epsilon $ (separately from $q $), etc. Mapping out the dependence just on $\gamma_E $, $q/\epsilon $ and $\kapcom $
took approximately 1500 nonlinear simulations at a total cost of around 4.5 million core hours. Adding even two or three more parameters to the search would take computing requirements beyond the limit of resources today, but not of the near future.

\begin{acknowledgments}
We are grateful for helpful discussions with I. Abel,  G. Colyer, R. Kerswell and A. Zocco. This work was supported by STFC (AAS), the Leverhulme Network for Magnetised Plasma Turbulence and the Wolfgang Pauli Institute, Vienna. Computing time was provided by HPC-FF and by EPSRC grant EP/H002081/1. 
\end{acknowledgments}

\bibliography{references}
\end{document}